\documentclass{ws-ijmpcs}

  \usepackage{amssymb}
   \usepackage{amsmath}
    \usepackage{epsfig}
     \usepackage{bm}
      \usepackage{pifont}

\def\Li{\relax\ifmmode{\textbf{Li}_{2}}\else{Li$_2${ }}\fi}

\newcommand{\half}{\frac{1}{2}}

\newcommand{\be}{\begin{equation}}
\newcommand{\ee}{\end{equation}}

\newcommand{\ba}{\begin{eqnarray}}
\newcommand{\ea}{\end{eqnarray}}
\newcommand{\bg}{\begin{gather}}
\newcommand{\foma}{\end{gather}}

\newcommand{\nn}{\nonumber}

\newcommand{\baa}{\begin{align}}
\newcommand{\eaa}{\end{align}}

\def\pd{\partial}
\def\pdlm{\partial_\mu}
\def\pdln{\partial_\nu}

\def\e{\epsilon}

\def\pd{\partial}

\def\<{\langle}
\def\>{\rangle}

\def\g{\gamma}  \def\G{\Gamma}
\def\d{\delta}  
\def\l{\lambda}   
\def\s{\sigma}
\def\r{\rho}

\def\m{\mu}
\def\n{\nu}

\def\({\left(}
\def\[{\left[}
\def\){\right)}
\def\]{\right]}

\def\pd{\partial}

\def\trimmarks{}

\begin{document}

\markboth{I.O. Cherednikov \& T. Mertens}
{Generalized Loop Space and Evolution of Wilson Loops}

%
\catchline{}{}{}{}{}
%

\title{Generalized Loop Space and Evolution of the Light-Like Wilson Loops}

\author{Igor O. Cherednikov}

\address{EDF, Departement Fysica, Universiteit Antwerpen,
Antwerp, B-2020 Belgium\\
igor.cherednikov@uantwerpen.be}

\author{Tom Mertens}
\address{EDF, Departement Fysica, Universiteit Antwerpen,
Antwerp, B-2020 Belgium\\
tom.mertens@uantwerpen.be}

\maketitle

\begin{history}
\end{history}

\begin{abstract}
Equations of motion for the light-like QCD Wilson loops are studied in the generalized loop space (GLS) setting. 
To this end, the classically conformal-invariant non-local variations of the cusped Wilson exponentials lying (partially) on the light-cone are formulated in terms of the 
Fr\'echet derivative. The rapidity and renormalization-group behaviour of the gauge-invariant quantum correlation functions 
(in particular, the three-dimensional parton densities) are demonstrated to be connected to certain geometrical properties of the Wilson loops defined in the GLS.
\keywords{Wilson lines and loops; generalised loop space; QCD factorization.}
\end{abstract}

\ccode{PACS numbers:13.60.Hb,13.85.Hd,13.87.Fh,13.88.+e}

\section{Introduction}
The QCD factorization approach to the analysis of the semi-inclusive high-energy processes entails the introduction of 
transverse-momentum dependent parton densities (TMD), which generalise the collinear (integrated) PDFs and contain essential information about three-dimensional intrinsic structure of the nucleon [\refcite{TMD_basic}].
In Ref. [\refcite{ICh_FBS_2014}] the following factorization scheme (valid in the large Bjorken-$x$ regime) for a generic {\it transverse-distance dependent} quark distribution function 
\begin{equation}
{\cal F} \left(x, {\bm b}_\perp\right)
=
\int\! d^2 k_\perp \  {\rm e}^{- i b_\perp k_\perp}  {\cal F} \left(x, {\bm k}_\perp\right)
\label{eq:TDD_def}
\end{equation}
has been proposed 
\begin{equation}
{\cal F} \left(x, {\bm b}_\perp; \eta, \mu^2 \right)
\approx
{\cal H} (\mu^2, P^2) \cdot {\Phi} (x, {\bm b}_\perp; \eta, \mu^2 ) ,
\label{eq:LargeX_factor}
\end{equation}
where the $x$-independent jet function ${\cal H}$ describes the incoming (collinear) partons and
the soft function $\Phi$ can be defined as the Fourier transform of an element of the {\it generalized loop space}
\begin{equation}
{\Phi} (x, {\bm b}_\perp; \eta, \mu^2 )
= \int\!dz^- \ {\rm e}^{-i (1-x) P^+ z^-} \ {\cal W}_* (z^-, {\bm b}_\perp; \eta, \mu^2 ) \ .
\label{eq:soft_LargeX}
\end{equation}
Here the $\eta$ and $\mu$ stand for the rapidity and ultra-violet regulators, respectively. 
The Wilson loop ${\cal W}_{*}$ reads  
\begin{equation}
{\cal W}_{*} (z^-, {\bm b}_\perp; \eta, \mu^2 ) =
 \langle 0 | \  {\cal U}_P^\dag [\infty; z] {\cal U}_{n^-}^\dag[z; \infty] {\cal U}_{n^-} [\infty ; 0] {\cal U}_P [0; \infty]
  \  | 0 \rangle , 
\end{equation}
and appears to be a cusped configuration consisting of two off-light-cone ${\cal U}_P$ with $P^2 \neq 0$, and two light-like ${\cal U}_{n^-}$ with $(n^-)^2 = 0 $ Wilson lines.
The quark jet effects in ${\cal H}$, therefore, are separated from  the soft function $\Phi$, which accumulates information about the intrinsic $3D$-structure of the nucleon in the large-$x$ domain, the latter being available at the planned EIC and Jefferson Lab, see Ref. [\refcite{EXP_Rev}] and Refs. therein.

As it follows from the factorization formula, Eq. (\ref{eq:LargeX_factor}), the soft function $\Phi$ contains the {\it rapidity} as well as {\it ultraviolet singularities} of the TMD distribution (\ref{eq:TDD_def}). The complex structure of the UV and light-cone (rapidity) divergences and their crucial effects on the evolution of TMDs with the emphasis on the gauge invariance and the properties of the anomalous dimensions has been studied in detail in Refs. [\refcite{CS_basic}]. 
On the other hand, as an element of the GLS, the Wilson loop ${\cal W}_{*}$ obeys the integro-differential equations of motion, which prescribe the behaviour of the quantum correlation functions containing ${\cal W}_{*}$ with respect to the shape variations of the underlying paths [\refcite{GLS}]. 
Let us show that the connection between certain (diffeomorphism-invariant) transformations in the GLS  and classically conformal invariant shape transformations allows one to simplify the calculation of the evolution kernels for the TMD (\ref{eq:TDD_def}). In particular, the rapidity differential operators can be represented in terms of the  {\it Fr\'echet differentials} enabling the derivation of the full set of the evolution equations. 

\begin{figure}[h]
			\includegraphics[width=1.\textwidth]{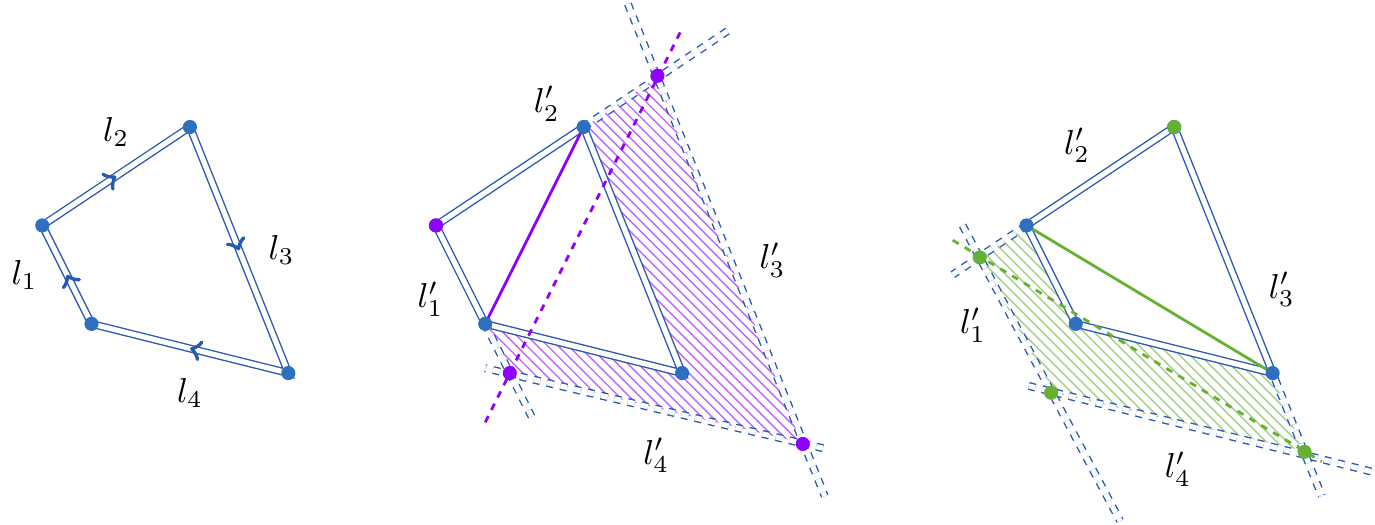}
			\caption{Quadrilateral light-like Wilson loop  (left panel) and the examples of the shape 
			transformations generated by the operators (\ref{eq:newdiffop}) (right panels).}
			\label{fig:WLQ_possible_area_var_LC}
		\end{figure}

To this end, let us consider the generic quadrilateral contour\footnote{The singularities and renormalization properties of the light-like Wilson polygons have been introduced and extensively studied in Refs. [\refcite{WL_LC_basic}].}, Fig. \ref{fig:WLQ_possible_area_var_LC}, with the sides given by the vectors
\begin{eqnarray}
  \ell_1^\mu & = &\ell_1 (1^+,0^-, \mathbf 0_\perp) , \   \ell_2^\mu = \ell_2(0^+,1^-, \mathbf 0_\perp), \nn \\   \ell_3^\mu & = & - \ell_1 (1^+,0^-, \mathbf 0_\perp) , \    \ell_4^\mu = - \ell_2 (0^+,1^-, \mathbf 0_\perp) . 
\end{eqnarray}
One can introduce a class of the path transformations generated by the differential operators [\refcite{WL_UA}]
		\begin{equation}\label{eq:newdiffop}
			S_{ij}\frac{\delta}{\delta S_{ij}}  =
			 (2{\ell}_i \cdot \ell_j) \frac{\partial}{\partial  (2\ell_i \cdot \ell_j)}  , \
			S_{ij}   =  (\ell_i + \ell_j)^2
			 \ ,		
			\end{equation}
			and
			\begin{equation}
			\Bigg\langle\frac{\delta}{\delta \ln S}\Bigg\rangle_1
			=
			S_{12}\frac{\delta}{\delta S_{12}} +  S_{23}\frac{\delta}{\delta S_{23}} , \  \Bigg\langle\frac{\delta}{\delta \ln S}\Bigg\rangle_2
			=
			S_{23}\frac{\delta}{\delta S_{23}} +  S_{34}\frac{\delta}{\delta S_{34}}, \text{etc.}
			\label{eq:newdiffop_1}
			\end{equation}
		
It is easy to see that the rapidities $y_{i}$ associated with the light-like vectors $\ell_{i}$, being formally infinite, can be regulated as follows:
\begin{equation}
 y_{1}
 =
 \frac{1}{2} \ \ln \frac{\ell_{1}^+}{\ell_{1}^-}
 \sim  \pm \frac{1}{2}
\lim_{\eta_1 \to 0}  \ \ln \frac{(\ell_1 \cdot \ell_2)}{\eta_1} .
\label{eq:rapidity}
\end{equation}
Hence, the differential operators (\ref{eq:newdiffop_1}) are related to the logarithmic rapidity derivative via
\begin{equation}
\frac{d }{d \ln S_{ij}} \sim \pm \frac{d }{d y_{i}}.
\end{equation}
We conjecture then that the rapidity evolution of a correlation function with light-like cusped Wilson loops corresponds to a shape-transformation law of a specific class of elements of the GLS. To reveal this law, we address the shape-transformations in the GLS by means of the so-called {\it Fr\'echet derivative} [\refcite{GLS}].

By definition, the logarithmic Fr\'echet derivative associated with a given vector $V$ reads
\be\label{eq:frechetder1}
			D_V [{U}_\gamma]
			=
			{U}_\g \cdot \int\limits_0^1\! dt \ {U}_{\g^t}\cdot {\cal F}_{\m\n}(t) \[V^\m(t)\wedge \dot{\g}^\n(t) \] \cdot {U}_{\g^t}^{-1} ,
		\ee	
for the Wilson exponential ${U}_\gamma$ evaluated along a given trajectory $\gamma$, where
\be
	{U}_{\gamma^t} = {\cal P} \exp{\left[i g \int\limits_{0}^t\! {\cal A}_\m(x)\ \dot{\gamma}^\m d\sigma \right]_\gamma} \ , \ x_\mu (\sigma) = \dot{\gamma}_\mu \sigma \  ,  
	\label{eq:paralleltransporter_0}
\ee
\begin{equation}
\ \sigma \in [0,1] \ , \  x_\mu (0) = x_\mu (1)\ , \ {U}_{\g} = {U}_{\g^1}  .
\end{equation}

Define a vector field
\begin{equation}
V^\m = V_1^\m+V_2^\m = (\ell_1^{+} \ , \ \ell_2^{-} \ , \ \bm 0_\perp)
\label{eq:V_1}
\end{equation}
which generates the angle-conserving transformations, Fig. \ref{fig:WLQ_possible_area_var_LC}. 
Let us show that the operator (\ref{eq:newdiffop_1}) transforms our ${\cal W}_*$ equivalently to this differential if the trajectory $\gamma$ is chosen as in Fig. \ref{fig:WLQ_possible_area_var_LC} and the generating vector $V$ defined in Eq. (\ref{eq:V_1})
\be
\left(S_{12}\frac{ \delta}{\delta S_{12}} + S_{23}\frac{ \delta}{\delta S_{23}}\right) \  {\cal W}_*  = D_{V} \ {\cal W}_*  ,  \  		 
{\cal W}_{*}
=
\Big\langle 0 \Big| \frac{1}{N_c} {\rm Tr} \ {U}_* \Big| 0 \Big\rangle .
\ee

\section{Calculation of the leading-order contributions}
\label{sec:frechet_lo}
Expand Eq. (\ref{eq:frechetder1}) to the leading non-trivial order:
		\ba
		 & &	D_V[{\cal W}_\g]^{(1)} = \nn \\
		 & &\oint\limits_0^1dt\  \left[\left(
				\oint\limits_0^t {\cal A}_\s (x(s)) \ \frac{dx^\s}{ds} ds \cdot
				\left\lbrace\pdlm {\cal A}_\n(y(t))- \pdln {\cal A}_\m (y(t))\right\rbrace \left\lbrace V^\m(y(t))\wedge \dot{\g}^\n(y(t))\right\rbrace
				\right)\right.\nn\\
				&&\ \ \ \ \
				\left.
				- \left(
				\left\lbrace\pdlm {\cal A}_\n (y(t))- \pdln {\cal A}_\m (y(t))\right\rbrace \left\lbrace V^\m(y(t))\wedge \dot{\g}^\n(y(t))\right\rbrace 
				\oint\limits_0^t {\cal A}_\l (x(u)) \ \frac{dx^\l}{du} du
				\right)
				\right]\nn\\
				&&\ \ \ \ \
				+\oint\limits_0^1 {\cal A}_\s (x) \frac{dx^\s}{ds} ds  \oint\limits_0^1 dt 
				\left\lbrace\pdlm {\cal A}_\n (y(t))- \pdln {\cal A}_\m (y(t))\right\rbrace \left\lbrace V^\m(y(t))\wedge \dot{\g}^\n(y(t))\right\rbrace
				 . \nn\\
		\label{eq:LO_1}
		\ea
We assume that the gluon propagator in the Feynman gauge reads in the coordinate space
		\be
			\langle 0 | T [A_\mu^a (x)A_\nu^b (y) ] | 0 \rangle
=			
			 D_{\m\n}^{ab}(x-y)
			=
			K_\epsilon \ \frac{g_{\m\n}\d^{ab}}{\left[-(x-y)^2\right]^{1-\e}}  ,
		\ee
	where 
	\begin{equation}
	K_\epsilon = \frac{(\m^2 \pi)^\e}{4\pi^2}\G(1-\e). 
	\end{equation} 

	Let us consider first the generating vector 
	\begin{equation}
	V_1^\m = ({\ell}_1^{+} \sigma, 0^{-}, \mathbf 0_{\perp}) \ , \  \sigma \in [0,1] .
	\end{equation}
Computation for the vector
	\be
	V_2^\m = (0^+, {\ell}_2^{-} \sigma', \mathbf 0_{\perp})
	\ee 
	runs similarly. 
The contributions from the wedge product 
\be 
V_1^\m(y(\sigma))\wedge \dot{\g}^\n(y(\sigma)
\ee 
can be described as follows:
		\begin{itemize}
\label{list1}

			\item {The sides ${\ell}_1$: $V^\mu \wedge \dot{\g}^\nu =0 $ and ${\ell}_3$: $V^\mu \wedge \dot{\g}^\nu =0 $, by the asymmetry of the wedge product and the fact that the vectors are parallel;}

			\item {The side ${\ell}_2$: $V^\mu \wedge \dot{\g}^\nu = -\ell_1^{+} \ell_2^{-}(\pd_{+}\wedge \pd_{-})$, by the (anti-)linearity of the wedge product;}


			\item {The side ${\ell}_4$: $V^\m \wedge \dot{\g}^\n =0 $, since the vector field equals zero along this part of the path.}

		\end{itemize}
We have, therefore, the following combinations of the gluon propagators to be evaluated:
	
		\subsection{$\pdlm D_{\s\n}(x-y)-\pdln D_{\s\m}(x-y)$ term with $x\in \ell_1$}
		
	Given that 
				\ba
					x &=& \sigma \ell_1 ,\ \sigma \in [0,1] \\
					y &=& \ell_1 + \sigma' \ell_2,\ \sigma'\in [0,1] \ ,
				\ea
	one gets
				\ba
					dx^\s &=& \left(\frac{dx^\s}{d\sigma}\right)d\sigma=(\ell_1^{+}, 0^{-}, \mathbf 0_\perp)d\sigma\nn\\
			dy^\n &=& \left(\frac{dy^\n}{d\sigma'}\right)d\sigma'=(0^{+},\ell_2^{-}, \mathbf 0_\perp)d\sigma'=\dot{\g}(\sigma')d\sigma'\nn\\
					x-y &=& (\sigma - 1) \ell_1 - \sigma' \ell_2\nn\\
					(x-y)^2 &=& -2(\sigma-1)\sigma' \ (\ell_1^{+}\ell_2^{-}) \ .\nn
				\ea
			
Straightforward computation yields	

				\ba\label{goal1}
					&&\int\limits_0^1\ d\s'\ d\s\frac{dx^\r}{d\s}
						\left(\frac{\pd}{\pd y^\m} D_{\r\n}(x-y) - \frac{\pd}{\pd y^\n} D_{\r\m}(x-y)\right)
						\left[V^\m(y) \wedge \dot{\g}^\n(y)\right]\nn\\
										&=& 	
							\half K_\e \frac{S_{12}^\e}{\e}\label{result1_4_1} .
				\ea
It is worth noticing that the same result can be obtained by applying the derivative $\ell_1\frac{ \partial}{\partial \ell_1}$ to the
			original integral
				\be
					\ell_1 \frac{\partial}{\partial \ell_1}K_\e \oint\! \frac{g_{\m\n}\ dx^\m\ dy^\n}{\left(-(x-y)^2\right)^{1-\e}}
					=
					\ell_1 \frac{\partial }{\partial \ell_1}K_\e \oint\! \frac{(\ell_1 \ell_2)\ d\s\ d\s'}{\left(-(2\ell_1 \ell_2(\s - 1) \s')^2\right)^{1-\e}} =
					\half K_\e \frac{S_{12}^\e}{\e} .
				\ee
		
		\subsection{$\pdlm D_{\r\n}(x-y)-\pdln D_{\r\m}(x-y)$ term with $x\in \ell_2$}
			
			This term is trivially zero since it represents the self-energy of a light-like WIlson line.
		
		\subsection{$\pdlm D_{\r\n}(x-y)-\pdln D_{\r\m}(x-y)$ term with $x\in \ell_3$}	
			
		One has
				\be
					\int\limits_0^1\ d\s'\ d\s\frac{dx^\r}{d\s}
						\left(\frac{\pd}{\pd y^\m} D_{\r\n}(x-y) - \frac{\pd}{\pd y^\n} D_{\r\m}(x-y)\right)
						\left[V^\m(y) \wedge \dot{\g}^\n(y)\right] = 0 .
				\ee
that is equal to the result of the differentiation $\ell_1\frac{ \partial }{\partial  \ell_1}$.
		
		\subsection{$\pdlm D_{\r\n}(x-y)-\pdln D_{\r\m}(x-y)$ term with $x\in \ell_4$}	
	We introduce the parametrization
				\ba
					x &=& -(1-\s) \ell_4 ,\ \sigma \in [0,1] \ , \\
					y &=& \ell_1 + \s' \ell_2 \ ,
					\ \sigma'
				\in
				[0,1] 
				\ea
and split up the calculations into the two terms
$\pdlm D_{\r\n}(x-y)$ and $-\pdln D_{\r\m}(x-y)$.
The first term $\pdlm D_{\r\n}(x-y)$ then returns
				\ba\label{eq:result2}
					& & \int\limits_0^1\ d\s'\ d\s\frac{dx^\r}{d\s}
						\left(\frac{\pd}{\pd y^\m} D_{\r\n}(x-y)\right)
						\left[V^\m(y) \wedge \dot{\g}^\n(y)\right]
						=\nn\\
							&&-2 (\epsilon -1) \int\limits_0^1\ d\s'\ d\s\ \left[\ell_1\cdot (\ell_1 + \s' \ell_2 + (1-\s) \ell_4) \right]\frac{(\ell_2\cdot \ell_4)}
							{\left(-(\ell_1+\s'\ell_2+(1 - \s)\ell_4)^2\right)^{2-\e}} ,\nn\\
				\ea
		while the second term can be shown to give zero [\refcite{CM_2014}].

The same procedure applies to the generating vector  
\be
V_2^\mu = (0^{+}, \ell_2^{-},\mathbf 0_\perp)
\ee
 with
			the point $y$ being attached to the side $\ell_3$.

Therefore, we obtain
				\be
					\Bigg\langle\frac{\delta}{\delta \ln S}\Bigg\rangle_1\  {\cal W}_* 
			=
			\left( S_{12}\frac{\delta}{\delta S_{12}} +  S_{23}\frac{\delta}{\delta S_{23}} \right)  \  {\cal W}_*  = D_V \ {\cal W}_* ,		
 		\ee
				with 
		\be 
		V^\m = V_1^\m+V_2^\m = (\ell_1^{+} \ , \ \ell_2^{-} \ , \ \mathbf 0_\perp) .
		\ee
				Taking into account the renormalization properties of the light-like Wilson quadrilateral loop [\refcite{WL_LC_basic}], we conclude that
\begin{equation}
 \mu \frac{d}{d\mu} \ \left[ D_V \  {\cal W}_* \right]
 =
- \sum \Gamma_{\rm cusp} ,
\label{eq:final}
\end{equation}
where $\Gamma_{\rm cusp}$ denotes the light-cone cusp anomalous dimension [\refcite{WL_LC_basic,Gamma_cusp}] and the summation over the number of cusps is assumed.

\section{Conclusions}

To summarise:

\begin{itemize}

\item 
The logarithmic Fr\'echet derivative interpreted as a diffeomorphism-induced differential operator associated with a generating vector field $V^\m$ is shown to be equivalent to the non-local infinitesimal shape-transformation introduced in Ref. [\refcite{WL_UA}].

\item
Therefore, a specific class of the motions in the GLS, referred as the classically conformal-invariant transformation, can be introduced in terms of the Fr\'echet derivative. Because diffeomorphisms do not produce new cusps, the number of cusps is diffeomorphism-invariant. We conjecture that the light-like cusped Wilson loops possessing different number of cusps correspond to different physical objects obeying different evolution laws.

\item
Application of the developed formalism to the derivation of the evolution equations for the TMDs on the light-cone is a subject of ongoing investigation and will be reported elsewhere. 

\end{itemize}


\section{Acknowledgements}
We thank Frederik Van der Veken and Pieter Taels for long-term fruitful collaboration and inspiring discussions.



  \end{document}